\documentstyle[mnextra]{mn}

\nobrackets

\def\lap{\mathrel {\hbox{\rlap{\hbox{\lower4pt\hbox{$\sim$}}}\hbox{$<$}}}}
\def\gap{\mathrel {\hbox{\rlap{\hbox{\lower4pt\hbox{$\sim$}}}\hbox{$>$}}}}
\def\lsim{\mathrel{\hbox{\rlap{\hbox{\lower4pt\hbox{$\sim$}}}\hbox{$<$}}}}
\def\gsim{\mathrel{\hbox{\rlap{\hbox{\lower4pt\hbox{$\sim$}}}\hbox{$>$}}}}
\def\Mo{{\rm M_\odot}}
\def\hMo{h^{-1}\; {\rm M_\odot}}
\def\hmpc{h^{-1}\;{\rm Mpc}}
\def\hkpc{h^{-1}\;{\rm kpc}}
\def\etal{et al.\ }
\def\ra{R_{\rm A}}
\def\dmr{DMR-normalized\ }
\def\cobe{$COBE$\ }
\def\cdmr{$COBE$-DMR\ }
\def\OmegaB{\Omega_{\rm B}}
\def\sigmag{{ \sigma_{8,{\rm gal}}}}
\input{epsf}

\begin{document}

\title[$COBE$-normalized CDM]{Large-scale structure in  $COBE$-normalized cold dark matter cosmogonies}


\author[Cole et al]{Shaun Cole${}^1$, David H. Weinberg${}^2$, Carlos S. 
Frenk${}^1$, and Bharat Ratra${}^{3,4}$ \\
$^1${Department of Physics, University of Durham, South Road, Durham DH1 3LE}\\
$^2${Department of Astronomy, Ohio State University, 174 W. 18th Avenue, 
Columbus, OH 43210, USA} \\
$^3${Center for Theoretical Physics, Massachusetts Institute 
of Technology, Cambridge, MA 02139, USA} \\
$^4${current address: Department of Physics, Kansas State 
University, Manhattan, KS 66506, USA} }

\maketitle

\begin{abstract}

We study the clustering of the mass distribution in cold dark matter models
using large cosmological N-body simulations. We investigate spatially-flat
models with a cosmological constant and scale-invariant ($n=1$) primordial
power spectra, as well as open-bubble inflation models. All the models
we consider are normalized according to the fluctuation amplitude measured 
in the \cdmr microwave background anisotropy data. With an age of the
universe $t_0\approx 14$ Gyr (12 Gyr) for the flat (open) models, a baryon
mass density parameter $\OmegaB = 0.0125 h^{-2}$, and a reasonable
assessment of the systematic uncertainties in the cumulative cluster mass
function, the observed abundance of rich galaxy clusters leads to tight
constraints on the mass density parameter $\Omega_0$. The allowable ranges
are $0.4\la\Omega_0\la 0.5$ for open models and $0.25\la\Omega_0\la 0.4$
for flat models. The upper limits on $\Omega_0$ can be relaxed if one
lowers the Hubble pararameter and increases the age of the universe, but
$h \lsim 0.25$ is required for $\Omega_0=1$ to be allowed.
The constraints also change if one allows tilted primordial power spectra.
An $\Omega_0=1$ cold dark matter model with $h=0.5$
can be constructed to satisfy both the cluster and DMR constraints, but 
it requires a tilted primordial power spectrum, with $n \approx 0.8$ and 
a corresponding 
contribution to the DMR signal from gravitational waves that reduces 
the implied $\sigma_8$ by a further 27\%. We compare the
evolved mass correlation functions and power spectra of the most promising
of our N-body models with those of galaxies in the APM survey. The flat
models have steep correlation functions at small scales and require the
galaxy distribution to be antibiased on scales $r\lsim 8 \hmpc$. The open
models require little or no antibias on small scales and a positive bias on
large scales; these biases are small for $\Omega\simeq 0.4$,
implying that, in this case, galaxies approximately trace the mass over a
wide range of scales.  The lack of a positive bias on small scales in
almost all of these N-body models is difficult to reconcile with the mean
mass-to-light ratio of cluster galaxies which, if $\Omega_0 \gsim 0.2$,
implies that galaxies are overabundant in clusters relative to the
field. The tilted $\Omega_0=1$ model, on the other hand, does require that
galaxies be positively biased on small scales, and that the bias to become
stronger on larger scales. We also compute the topology of isodensity
contours in these models, obtaining theoretical predictions that are less
sensitive to the details of galaxy formation.

\end{abstract}


\section{INTRODUCTION}

Earlier this century, Einstein and de Sitter (\shortcite{einstein}) 
suggested that until there 
was sufficient observational data, progress in constraining the 
cosmological model could perhaps be best made by focussing discussion
on the restricted model that now bears their names. In this Einstein-de 
Sitter model, the spatial hypersurfaces are flat, the cosmological constant 
$\Lambda$ vanishes, and the universe is dominated by pressureless matter,
so the mass density parameter $\Omega_0 = 8\pi
G\rho_b(t_0)/(3H_0{}^2)=1$, where $\rho_b(t_0)$ is the mean mass
density today, $H_0$ is the Hubble parameter, and $G$ is the Newtonian 
gravitational constant. 

In the 1980s, the ideas of inflation and pressureless, cold dark matter led to
a more encompassing and far more predictive version of the 
Einstein-de Sitter model: the cold dark matter (CDM) scenario
(\cite{peebles82,bfpr84,defw85}). In this picture, inflation explains why the 
universe is homogeneous on large scales, and 
quantum fluctuations generated during inflation provide a source of 
primordial perturbations that can grow through gravitational instability 
of the cold matter into the galaxies, clusters, and superclusters that are 
observed today.
The discovery of large-scale anisotropies in the cosmic microwave
background (CMB) by the \cdmr experiment (\cite{smoot92})
gave a big boost to the broad picture of structure formation by
gravitational instability since it revealed inhomogeneities in
the early universe at roughly the level predicted by theoretical models.
However, the DMR results also led to a powerful argument against
the simplest version of the cold dark matter model: when normalized to produce
the observed level of microwave background anisotropies, the model
predicted excessively massive galaxy clusters (e.g., Wright et al.
\shortcite{wright92}; Efstathiou, Bond, \& White \shortcite{ebw92}; 
Bahcall \& Cen \shortcite{bc93}; Bartlett \& Silk \shortcite{bs93};
White, Efstathiou \& Frenk \shortcite{wef93}). The level of discrepancy is 
only a factor of two or three in the fluctuation
amplitude, but it reinforces years of accumulating evidence for a difference
between the shape of the matter power spectrum predicted by standard
CDM and the shapes of the observed power spectra of galaxies and galaxy 
clusters (e.g., Bahcall \& Soneira \shortcite{bs83}; 
Klypin \& Kopylov \shortcite{kk83}; Maddox et al. \shortcite{apm90}; 
Efstathiou, Sutherland \& Maddox \shortcite{apm_nature}; Eke \etal
\shortcite{ecf96a}).

In the wake of the \cobe results, many variations on the cold dark matter 
theme have been explored.  Arguably, the most attractive of these models
are those that retain the assumptions of cold dark matter and an
inflationary origin of perturbations but drop the assumption that $\Omega_0=1$.
These low density CDM models come in two varieties, spatially flat models with 
a cosmological constant (e.g. Efstathiou, Sutherland \& Maddox
\shortcite{apm_nature}, \cite{os95}; 
Ganga, Ratra, \& Sugiyama \shortcite{ganga96c}; 
Liddle et al. \shortcite{llvw96}; 
\cite{lineweaver96,peacock96,ratra96}), 
and open models with $\Lambda=0$ 
(e.g., \cite{krss94,gorski95},\shortcite{gorski96b}; 
\cite{ganga96c,peacock96,ratra96}).  
One can consider models with both space curvature and non-zero $\Lambda$,
but the observations have not yet driven many theorists to such extremes.
Within some rather broad ranges of parameters, both flat and open models
can accommodate the DMR data on their own. A key element in tightening the 
range of acceptable parameters has again been the comparison between the 
DMR normalization and the predicted masses (or abundances) of galaxy clusters.
Most implementations of this approach have used semi-analytic
computations based on the Press-Schechter (\shortcite{ps74}) 
formalism (e.g., White et al. \shortcite{wef93}; \cite{borgani96};
Eke, Cole \&\ Frenk \shortcite{ecf96}; \cite{gorski96b};
Kitayama \& Suto \shortcite{kitayama96}; 
Viana \& Liddle \shortcite{vl96}).
In this paper we use large cosmological N-body simulations to
study the cluster mass function in DMR-normalized, open and flat CDM models
with $\Omega_0 = 0.1$, 0.2, 0.3, 0.4, 0.5, and 1. We also compute clustering 
properties of the non-linear mass distribution in the most promising of these 
models and in a tilted ($n<1$) $\Omega_0=1$ model.

Variants of the CDM scenario that retain the assumption of $\Omega_0 = 1$
are still under discussion. These variants include the tilted model which
we consider, but also may involve a broken 
primeval fluctuation spectrum, a gravitational wave contribution to the
CMB anisotropies observed by $COBE$-DMR, an 
admixture of hot dark matter, a high baryon fraction, a decaying particle 
that boosts the neutrino background, or a low Hubble parameter --- in short, 
one can fiddle with inflation, fiddle with the matter content, or fiddle with 
$H_0$ (see, e.g., \cite{be91,ss92,wvls96}).
These models can, to varying degrees,
account for the \cdmr observations and large scale structure,
provided that optically bright galaxies are biased with respect
to mass by a factor $b \sim 1.5-2$.  However, there are two other
arguments that generically favour low density models.  First is
the combination of recent estimates of the Hubble parameter,
which favour $h \equiv H_0/(100\;$km s$^{-1}$ Mpc$^{-1})>0.55$
(e.g., \cite{tanvir95}; Bureau, Mould \& Stavely-Smith \shortcite{bms96};
Sandage et al. \shortcite{sandage96}; 
Riess, Press \&\ Kirshner \shortcite{riess96}; \cite{giovanelli96}),
with estimates of globular cluster that 
imply an age of the universe $t_0 \gap 12\;$Gyr (\cite{cdkk96};
Salaris, Degl'Innocenti \& Weiss \shortcite{salaris96}; \cite{renzini96}). 
These estimates are difficult to reconcile if $\Omega_0 = 1$.  Second is the 
high baryon fraction in rich clusters of galaxies (e.g., \cite{henry79}; 
Henriksen \& Mushotzky \shortcite{hm85}); this is incompatible with 
$\Omega_0 = 1$ if the ratio of baryons to dark matter in clusters is equal to 
the universal value and constraints on the baryon density derived from 
standard nucleosynthesis theory and the observed light element abundances are 
correct (\cite{wnef93,markevitch96}; Buote \&\ Canizares \shortcite{bc96}; 
\cite{lubin96}).

Traditional inflation models lead to a spatially flat universe,
and hence to $\Omega_0 = 1$ if the cosmological constant vanishes.
The $\Lambda$-CDM models are perfectly compatible with standard
inflation if one considers a non-zero cosmological constant to be
physically reasonable. Open-bubble inflation models (Ratra \& Peebles 
\shortcite{rp94},\shortcite{rp95}; 
Bucher, Goldhaber \& Turok \shortcite{bgt95}; 
Yamamoto, Sasaki \& Tanaka \shortcite{yst95}) are variants of the scenario 
suggested by Gott (\shortcite{gott82}) and Guth \&\ Weinberg 
(\shortcite{gw83}), in which a spatially-open, inflating bubble nucleates 
via quantum tunneling within a spatially flat, exponentially expanding 
(inflating), de Sitter spacetime. The post-nucleation inflation epoch 
stretches the open bubble to encompass the observable part of the universe, 
and the quantum fluctuations of the inflaton field inside this spatially-open 
bubble provide the seeds for structure formation.

In the next section we describe our selection of models, with particular
attention to the normalization of the linear power spectrum.
We also describe our numerical simulation parameters.
In Section 3 we compare the mass function of clusters in these simulations
to observations.  The models that appear viable after this comparison
are the flat models with $\Omega_0 = 0.3$ or 0.4, and the open models with
$\Omega_0 = 0.4$ or 0.5.  An alternative viable model is  $\Omega_0=1$ 
with a tilted ($n<1$) primordial power spectrum. 
In Section 4 we present some statistical measures
of the clustering of the nonlinear mass distributions in these models:
the correlation function, the power spectrum, and the topology of isodensity 
surfaces.

These analyses do not include any 
effects of biased or anti-biased galaxy formation, but they give a sense of 
what sort of biases are needed to reconcile the model simulations with 
observations. In Section 5 we summarize our results and discuss the prospects for
distinguishing these models with future observations.

\section{Models}

\subsection{Parameter choices and normalization}

The primary parameter in the two classes of models we consider is
the density parameter $\Omega_0$.  However, the values of $h$ and $\OmegaB$
also have an influence on the shape of the matter power
spectrum, and through this on the amplitude of mass fluctuations for
the \cdmr normalization.  In light of recent evidence that tends to 
favour high values of $h$, we have decided to choose for each open model the
value of $h$ that gives $t_0 \approx 12\;$Gyr, i.e., the largest value that 
leaves the model marginally compatible with standard globular cluster age 
estimates.\footnote{
Note that lower values of $t_0$ might not be unreasonable. 
For example, Alcock et al. (\shortcite{alcock96}) 
argue for $t_0 = 12 \pm 1.5\;$Gyr on the basis of a revised RR Lyrae
distance scale.  A smaller $t_0$ allows for a larger $h$, 
hence more small-scale power and so lower $\Omega_0$.} 
For each flat-$\Lambda$ model we choose the value of $h$ that gives 
$t_0 \approx 14\;$Gyr.  For $\Omega_0=1$ models we take $h=0.5$.
Although our simulations do not include a separately 
treated baryon component, the value of the baryon fraction has a modest 
influence (weaker than the influence of $h$) on the shape and normalization 
of the power spectrum. To fix the initial amplitude and shape of the matter
power spectrum, we adopt $\OmegaB=0.0125h^{-2}$, the value advocated by 
Walker et al.\ (\shortcite{walker91}) on the basis of the observed light 
element abundances and standard nucleosynthesis. Recent estimates of the 
deuterium abundance in high-redshift Lyman-limit absorbers suggest that this 
value may be a factor of two high 
(\cite{songaila94,carswell94};  Rugers \& Hogan 
\shortcite{rh96a},b; see however Tytler, Burles \& Kirkman
\shortcite{tytler96b}) or low (Tytler, Fan \& Burles \shortcite{tytler96}; 
\cite{bt96}; see however \cite{wampler96}; Songaila, Wampler \& Cowie 
\shortcite{songaila96}).

On the scales modelled by the simulations in this paper, which are much 
smaller than present values of the Hubble or curvature radius, the shape 
of the energy-density perturbation power spectrum is 
quite accurately determined by taking account of the usual effects of matter
and radiation in processing the primordial 
energy-density perturbation power spectrum. 
We adopt the Bardeen et al. (\shortcite{bbks})
formula for the matter power spectrum, 
\begin{eqnarray}
   P(k) &\propto& \frac{k^n}  
   {\left[1 + 3.89q + (16.1q)^{2} + (5.46q)^3 + (6.71q)^4 \right]^{1/2}} 
\nonumber \\
&& \times \frac{\left[\ln(1+2.34q)\right]^2}{(2.34q)^2},
\end{eqnarray}
where $q=(k/\Gamma)h \, {\rm Mpc}^{-1}$.
The shape parameter $\Gamma$ is
\begin{equation}
   \Gamma = \Omega_0 h \,{\rm exp}(-\OmegaB-\OmegaB/\Omega_0)
\end{equation}
(\cite{sugiyama95}).  The slope of the primordial power spectrum, $n$,
we take to have the scale-invariant value ($n=1$),
except for the case of our one tilted model which has
$n \simeq 0.8$.  
This smooth analytic fit to the matter power spectrum
is fairly accurate in the limit $\OmegaB/\Omega_0 \ll 1$, but it does
not reproduce the wiggles in the power spectrum caused by oscillations
in the photon-baryon fluid when the baryon fraction is significant
(\cite{sugiyama95}).  We expect these wiggles to have minimal effect
on cluster masses or small scale clustering of the mass distribution,
especially for the relatively high values of $\Omega_0$ that are
favoured by our analysis in Section 3.

\begin{table*}
\centering
\caption{Simulation Parameters}
\begin{center}
\begin{tabular}{llllllllllllllllr} \hline
\multicolumn{1}{l} {Model} &
\multicolumn{1}{l} {$\Omega_0$ } &
\multicolumn{1}{l} {$\Lambda_0$  } &
\multicolumn{1}{l} {$h$} &
\multicolumn{1}{l} {$t_0/{\rm Gyr}$ } &
\multicolumn{1}{l} {$\Omega_{\rm B}$ } &
\multicolumn{1}{l} {$\Gamma$  } &
\multicolumn{1}{l} {$\sigma_8$  } &
\multicolumn{1}{l} {$a^{\rm i}$   } &
\multicolumn{1}{l} {$N_{\rm steps}$} \\
\hline 
O1   & 0.1 & 0.0 & 0.75 & 11.7& 0.022& 0.05& 0.1 & 0.98      &   2    \\
O2   & 0.2 & 0.0 & 0.70 & 11.8& 0.026& 0.12& 0.3 & 0.28      &  35    \\
O3   & 0.3 & 0.0 & 0.65 & 12.2& 0.030& 0.17& 0.5 & 0.15      &  93    \\
O4   & 0.4 & 0.0 & 0.65 & 11.7& 0.030& 0.23& 0.75& 0.1       & 168    \\
O5   & 0.5 & 0.0 & 0.6  & 12.3& 0.035& 0.27& 0.9 & 0.08      & 254    \\
L1   & 0.1 & 0.9 & 0.9  & 13.9& 0.035& 0.07& 0.7 & 0.15      & 150    \\
L2   & 0.2 & 0.8 & 0.75 & 14.0& 0.022& 0.13& 0.9 & 0.12      & 220    \\
L3   & 0.3 & 0.7 & 0.65 & 14.5& 0.030& 0.17& 1.05& 0.101     & 266    \\
L4   & 0.4 & 0.6 & 0.6  & 14.5& 0.035& 0.21& 1.1 & 0.09      & 275    \\
L5   & 0.5 & 0.5 & 0.6  & 13.5& 0.035& 0.27& 1.3 & 0.07      & 331    \\
E1   & 1.0 & 0.0 & 0.5  & 13.1&  --- & 0.5 & 1.35& 0.0605    & 327    \\
E2   & 1.0 & 0.0 & 0.5  & 13.1& 0.05 & --- & 0.55& 0.2       & 200    \\
\hline
\end{tabular} 
\label{tab:models}
\end{center}
\end{table*}

The normalization of the matter power spectrum makes use of spectral 
information on the very large scales probed by the \cdmr experiment.
For the $\Omega_0 = 1$ and open models, we take our values of $\sigma_8$, 
the rms, linearly extrapolated mass fluctuation in spheres of radius 
$8h^{-1}$ Mpc, from Table 1 of G\'orski et al. (\shortcite{gorski95}), 
which is based on an
analysis of the DMR two-year galactic-frame 53 and 90 GHz sky maps, with 
the observed quadrupole anisotropy moment excluded from the analysis.
For the flat models with a cosmological constant, we take values of 
$\sigma_8$ from Table 2 of Ratra et al. (\shortcite{ratra96}), 
which is based on an 
analysis of the DMR two-year ecliptic-frame 31.5, 53, and 90 GHz sky maps,
with the observed quadrupole anisotropy moment excluded from the analysis
(\cite{bs95}). Similar values of $\sigma_8$ for the flat-$\Lambda$ 
models are obtained by Sugiyama (\shortcite{sugiyama95}) and 
Stompor, G\'orski \& Banday (\shortcite{stompor95}), and for the 
$\Omega_0 = 1$ model by Bunn, Scott \& White (\shortcite{bsw95}), 
Sugiyama (\shortcite{sugiyama95}), and Stompor et al. (\shortcite{stompor95}). 
For the flat-$\Lambda$ models these estimates assume a scale-invariant
primordial matter power spectrum (spectral index $n=1$ in $P(k) \propto k^n$),
as predicted by the simplest inflation models.
The simplest open-bubble inflation models, which we adopt for our
open CDM models in this paper, also produce matter power spectra
that have this scale-invariant form on scales much smaller
than the present Hubble length (\cite{rp94}). 
Except for the tilted $\Omega_0=1$ model, 
our normalizations ignore the possible influence of primordial
gravity waves or mild early reionization on the \cdmr\ anisotropies.

There are various ways to analyze the DMR two-year data that lead
to slightly different $\sigma_8$ normalizations for a specified
cosmological model --- one can, for example, use either the ecliptic-
or galactic-frame sky maps and either include or ignore the observed
quadrupole anisotropy moment (which is the multipole most sensitive to 
the Galaxy model) in the analysis (e.g., Stompor et al. \shortcite{stompor95}).
There is thus no unique central $\sigma_8$ value for a ``DMR-normalized''
cosmological model, but for our purposes the differences between these 
different methods are not very significant relative to the observational
uncertainties in cluster masses discussed below in Section 3.

After we had completed our simulations, results from the analyses of the \cdmr
four-year data (\cite{bennett96,gorski96a}, and references therein) became 
available. In addition to new data, these analyses incorporate more 
sophisticated approaches to removing Milky Way contamination. The net effect 
is a slight downward shift in central estimates of the power spectrum 
normalization for the models, mostly a consequence of the new (four-year DMR)
Milky Way exclusion cut (\cite{banday97}). Different combinations of 
anisotropy maps, treatment of the quadrupole, and corrections for Milky Way 
emission again lead to slightly different estimates of $\sigma_8$ 
(\cite{gorski96a},b), but relative to the average of these DMR four-year 
estimates our adopted normalizations are typically high by $6-12\%$, on the 
same order as the 1-$\sigma$ DMR four-year data uncertainty.
Depending on spectrum and parameter values, the DMR four-year data uncertainty 
(including all known systematic uncertainties) in the $\sigma_8$ normalization
is $\sim \pm (16-19)\%$ (2-$\sigma$) (\cite{gorski96b}). 

\begin{figure*}
\centering
\centerline{\epsfxsize= 6.0 truein \epsfbox[20 0 545 750]{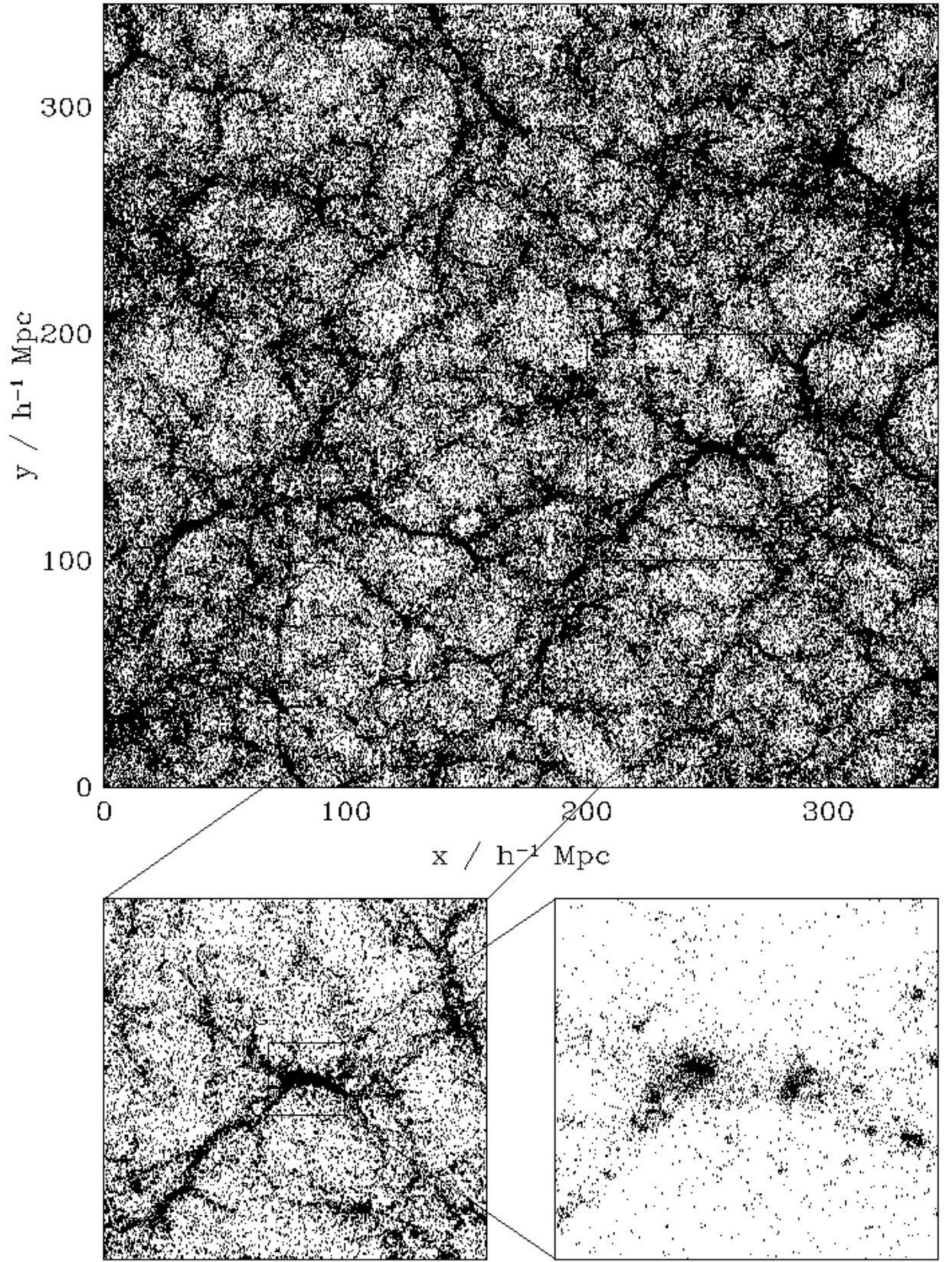}}
\caption{The evolved particle distribution in a $20\hmpc$ thick slice
of the $\Omega_0=0.4$, open universe simulation.  The large panel
shows the full cross section of the simulation box, while the two
lower panels show successively expanded views, $100\hmpc$ and $20\hmpc$
on a side respectively.  }
\label{fig:zoom}
\end{figure*}

\begin{figure*}
\centering
\centerline{
\epsfysize = 7.5 truein \epsfbox[90 40 520 740]{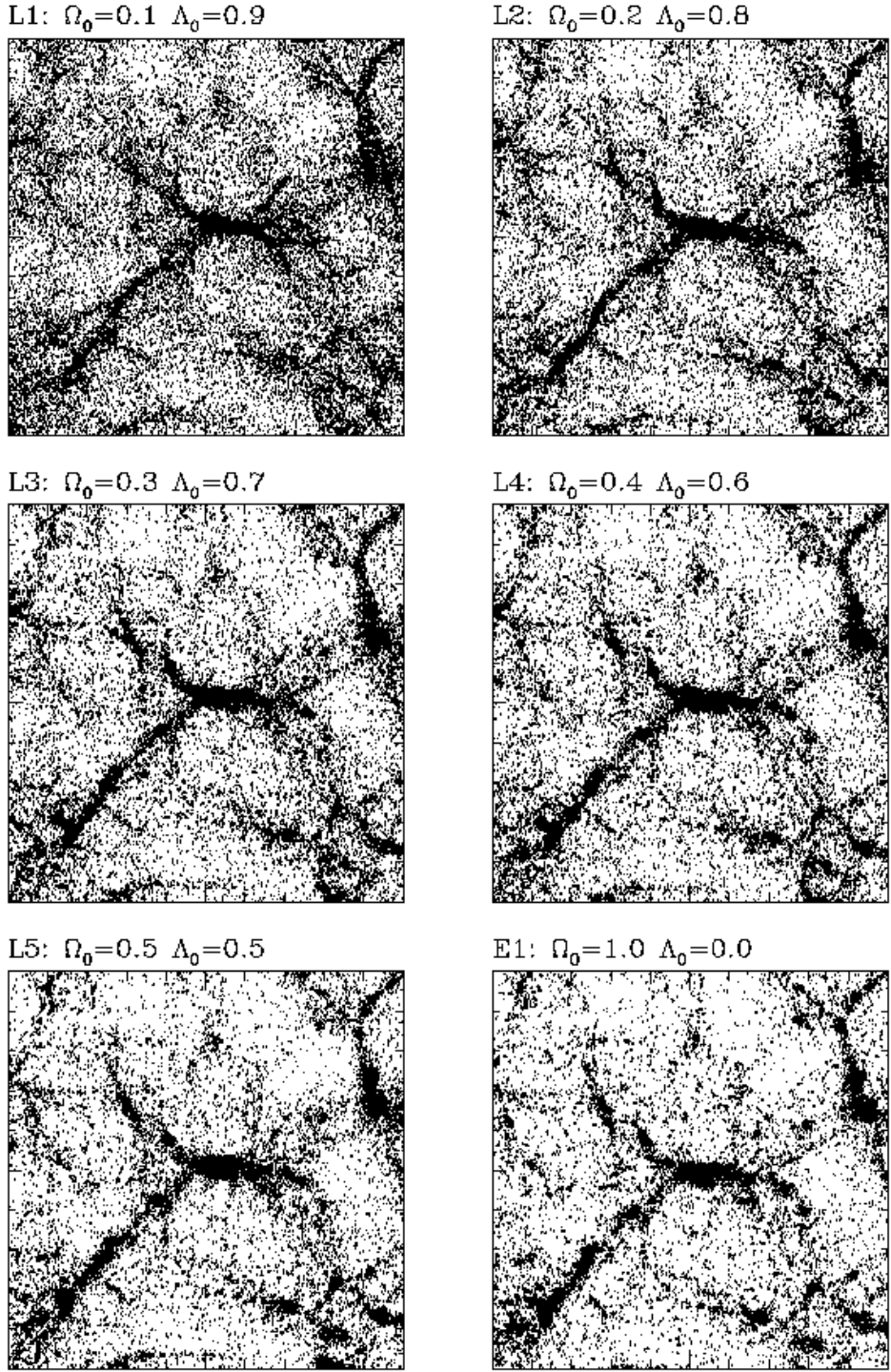}
}
\caption{Evolved particle distributions from the six DMR-normalized,
flat universe models.  The slices are $20\hmpc$ thick and $100\hmpc$
on a side, and they show the same portion of the simulation volume
that is shown in the lower left panel of Fig.~\ref{fig:zoom}.
 }
\label{fig:lslice}
\end{figure*}

\begin{figure*}
\centering
\centerline{
\epsfysize = 7.5 truein \epsfbox[90 40 520 740]{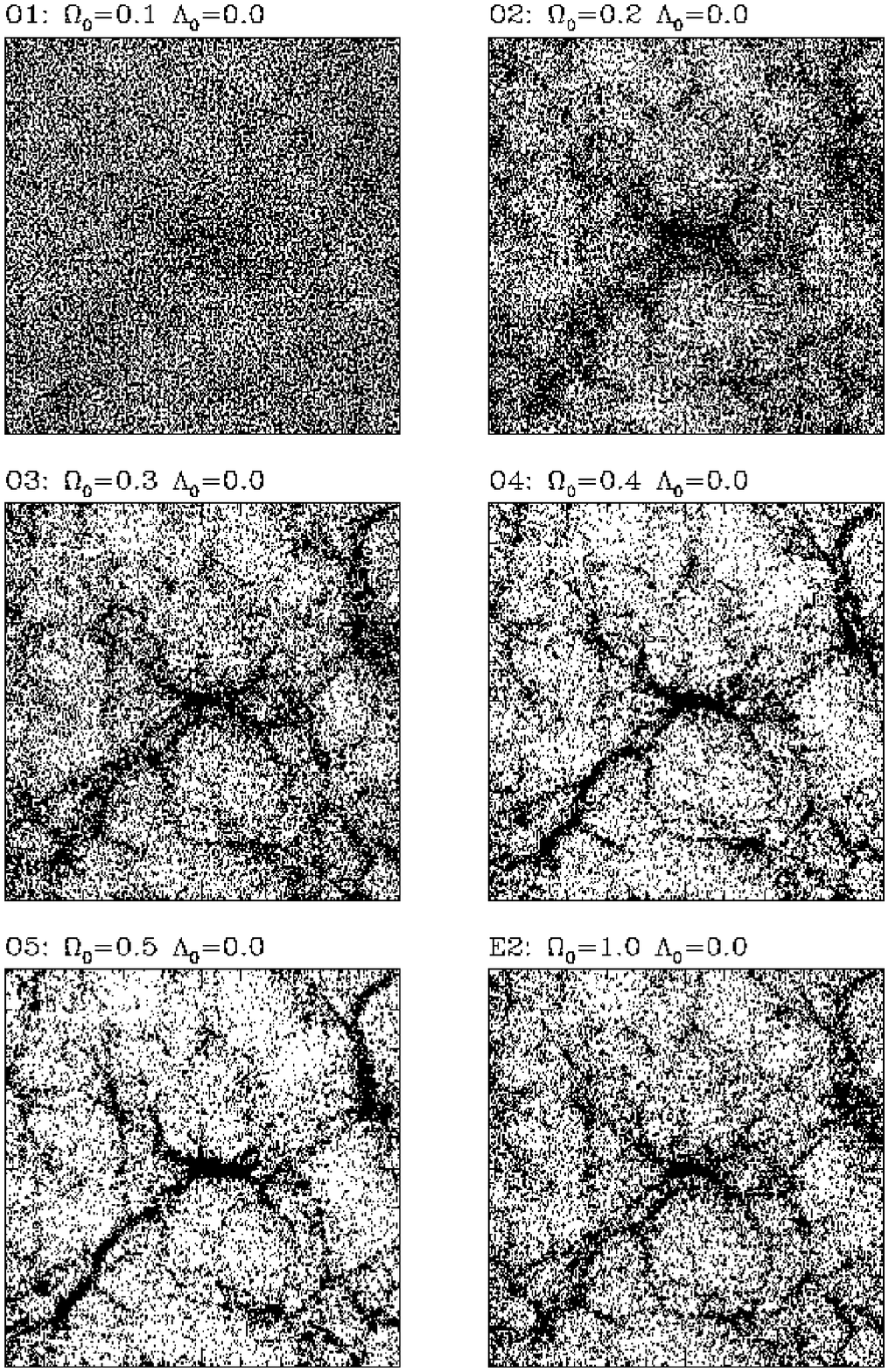}
}
\caption{Same as Fig.~\ref{fig:lslice} but for open universe models.
The bottom right panel shows the tilted $n=0.803$, $\Omega_0=1$ 
model, E2.
 }
\label{fig:oslice}
\end{figure*}

In addition to this series of models based on the simplest inflation
power spectra, we consider
an $\Omega_0=1$ model with a tilted primordial power spectrum, $n<1$. We have 
chosen the degree of tilt so that when the (power-law inflation model) 
gravity wave contribution to the DMR CMB fluctuations is included 
(\cite{bw96,llvw96}), the \dmr 
$\sigma_8=0.55$, which is about the value 
that White et al. (\shortcite{wef93}) conclude is needed to explain the 
mass function of galaxy clusters in an $\Omega_0 = 1$ universe.  This
prescription and a Hubble parameter of $h=0.5$ 
leads to $n=0.803$, for which the reduction in the implied $\sigma_8$ 
caused by the inclusion of gravitational waves is 27\%.

The values of $\Omega_0$, $\Lambda_0$ (in units of $3H_0^2$),
$h$, $t_0$, $\Omega_{\rm B}$, $\Gamma$, and $\sigma_8$ for all of our
models are listed in Table 1.  
We label the five flat-$\Lambda$ models L1--L5, the five open models
O1--O5, the DMR-normalized Einstein-de Sitter ($\Omega_0=1$) model E1,
and the tilted Einstein-de Sitter model E2. Note that for the model E1 
$\Gamma=0.5$, as is conventional in the standard CDM model,
so here we have effectively assumed $\Omega_{\rm B}=0$.

\subsection{Simulations}

Each of our simulations models a periodic cube of side 345.6$\hmpc$ (comoving).
We want large boxes so that we can get accurate statistical estimates of 
cluster abundances, and also because, in future work, we will use these 
simulations to create mock redshift catalogues of the Sloan Digital Sky Survey 
(see \cite{gw95}) and the Anglo-Australian 2-Degree Field Survey (see 
\cite{colless96}). We evolve the simulations using a modified version of Hugh 
Couchman's Adaptive Particle-Particle-Particle-Mesh (AP$^3$M, 
\cite{couchman91}) N-body code. The P$^3$M algorithm (\cite{edfw85}) 
allows us to simulate large volumes with
a force and mass resolution sufficient to yield accurate cluster masses.

Instead of starting our simulations from a regular particle grid, we use 
the technique described by 
Baugh, Gazta\~naga \& Efstathiou (\shortcite{baugh95}) 
and White (\shortcite{white95}) 
to generate ``glass'' initial 
conditions.  We lay down a Poisson distribution of $192^3$ particles, then
integrate these with the N-body code and sign-reversed gravitational
forces until they reach a state at which the gravitational forces
on all particles practically vanish.   With this approach, the initial particle
distribution is not regular, but random fluctuations in the particle
density field do not seed the growth of spurious structures.
For each simulation, we create a Gaussian random density 
field on a $192^3$ grid, using the same Fourier phases from one model
to the next but varying the power spectrum in accord with the model parameters.
We apply the Zel'dovich approximation to this density field
to compute displacements and peculiar velocities for the particles,
interpolating from the grid to the particle positions.
Simulations with glass initial conditions and grid initial conditions
yield similar statistical results once they are evolved into
the nonlinear regime (\cite{baugh95,white95}), 
but simulations with glass initial 
conditions do not retain a grid signature in uncollapsed regions.
We set the softening parameter of AP$^3$M's triangular-shaped cloud
force law to $\eta=270\hkpc$, $0.15$ of the grid spacing.
This choice corresponds approximately to a gravitational softening
length $\epsilon=\eta/3=90\hkpc$ for a Plummer force law,
and we adopt $\epsilon$ as our nominal force resolution.

The initial expansion factors of the simulations $a^{\rm i}$, 
listed in Table~\ref{tab:models}, were
determined by setting the amplitude of the initial power spectrum at 
the Nyquist frequency of the particle grid to be $0.3^2$ times that 
for an equivalent Poisson distribution of particles. Thus
$P_{\rm initial}(k_{\rm N}) = 0.3^2/\bar n$, where $\bar n$ is the
mean particle density and the Nyquist frequency is
$k_{\rm N}=\pi \bar n^{1/3} = (2 \pi/3.6)h\;{\rm Mpc}^{-1}$.
The residual power in the glass configuration is only 0.5\% of
that in a Poisson distribution at the Nyquist frequency
and drops very rapidly at longer wavelengths (see figure~A2 of
\cite{baugh95}). Thus this choice is safely in the regime
where the initial density fluctuations are large compared to
those present in the glass and yet in the linear regime
where the Zel'dovich  approximation remains accurate. In particular, 
no shell-crossing has occurred. (The O1 model has an amplitude lower than 
this choice at the final time. Consequently, we started this simulation
with fluctuation amplitude a factor of $2$ lower.)
The size of the time step $\Delta a$ was
chosen so that the following two constraints were satisfied throughout the
evolution  of the particle distribution. First, 
the rms displacement of particles in one timestep was less
than $\eta/4$. Second, the fastest moving particle moved less than
$\eta$ in one timestep. Initially these two constraints are comparable,
but at late times the latter constraint is more stringent, particularly
in low $\Omega_0$ simulations. We monitored energy conservation
using the Layzer-Irvine equation (equation 12b of \cite{edfw85}) 
and found that for this choice of timestep
energy conservation with a fractional accuracy of better than $0.3$\%
was achieved.
We also tested the inaccuracy incurred by these choices of starting 
amplitude and timestep by comparing the final particle positions with
two additional $\Omega_0=1$ simulations that were run starting from a 
fluctuation amplitude a factor of two lower and using time steps 
a factor of two smaller.
In each case we found the final particle positions to agree very accurately, 
with rms differences of less than $\epsilon$.
More importantly, the correlation functions
of each particle distribution were indistinguishable at scales
larger than $\epsilon=90 \hkpc$. Thus the statistical clustering
properties of these simulations have their resolution limited
by the particle mass and force softening 
and not by the choice of timestep or starting redshift.

The large panel of
Fig.~\ref{fig:zoom} shows a slice $20\hmpc$ thick (6\% of the 
simulation volume) through the evolved 
particle distribution of the $\Omega_0=0.4$, open universe model (O4).
The distribution follows
the familiar pattern of elongated, sharp, high density 
features and rounded voids that arises generically from the action of
gravitational instability on Gaussian initial conditions.
To illustrate the dynamic range of the simulation, we show 
expanded regions $100\hmpc$ and $20\hmpc$ on a side in the
lower two panels.   At the resolution level of the $20\hmpc$
panel, one sees that the coherent large scale features are 
composed mainly of dense, ellipsoidal clumps, connected by a
sprinkling of isolated particles.

Fig.~\ref{fig:lslice} shows $100\hmpc\times 100\hmpc\times
20\hmpc$ slices from our six DMR-normalized, flat universe models,
L1-L5 and E1.  Moving from $\Omega_0=0.1$ through successively higher
values of $\Omega_0$, we see a steadily increasing degree of clustering,
primarily because the value of $\sigma_8$ implied by DMR-normalization
is an increasing function of $\Omega_0$.  The mass per particle also
grows in proportion to $\Omega_0$, so the masses of the collapsed
structures increase more rapidly with $\Omega_0$.  The $\sigma_8$
normalization of the $\Omega_0=1$
model is only slightly larger than that of the $\Omega_0=0.5$ model, but
its structure is choppier because its initial conditions have more
small scale power ($\Gamma=0.5$ vs. $\Gamma=0.27$).

Fig.~\ref{fig:oslice} shows slices from the five open models, O1-O5,
and from the tilted $\Omega_0=1$ model, E2.  
We see the same trends as in Fig.~\ref{fig:lslice}, but here they
are much stronger, because in open models the $\sigma_8$ normalization
implied by the DMR data is more sensitive to the value of $\Omega_0$.  
The $\Omega_0=0.1$ model has virtually no collapsed structure on
scales that our simulation can resolve, and structure in the
$\Omega_0=0.2$ and $\Omega_0=0.3$ models is rather anemic.  
The structure in the E2 model is intermediate between that of the
O3 and O4 models, which have bracketing values of $\sigma_8$.

From these plots we can already anticipate the qualitative results for
cluster masses.  In each class of models, cluster masses will
increase with increasing $\Omega_0$ because of the greater level
of clustering and the higher mass density.  The dependence in the
open models will be stronger than in the flat models.

\begin{figure*}
\centering
\centerline{
\epsfxsize = 3.5 truein \epsfbox[80 165 430 725]{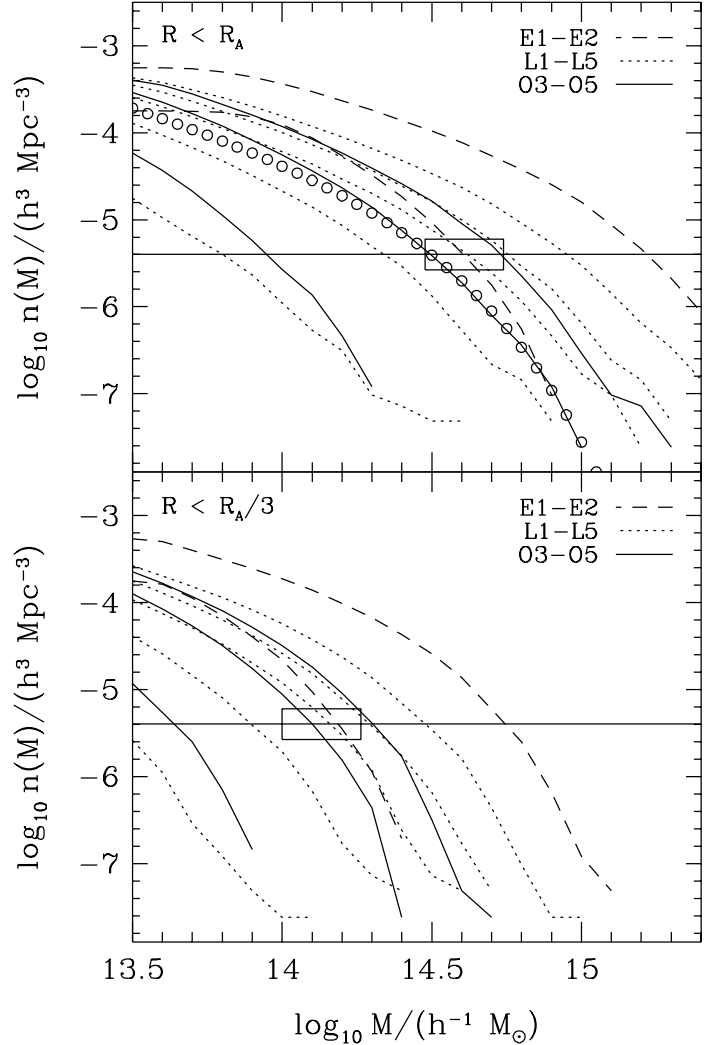}
}
\caption{Cumulative cluster mass functions; $n(M)$ is the number
density of clusters whose mass exceeds $M$.  In the upper panel,
masses are defined within the Abell radius $R_A=1.5\hmpc$.  
In the lower panel, masses are defined within $R_A/3$.
In each panel, the upper dashed line represents the
\dmr fiducial $\Omega_0=1$ model, E1, and the lower dashed line
the \dmr tilted $\Omega_0=1$ model, E2.
Dotted lines represent the $\Lambda$-dominated models L1-L5,
with $\Omega_0$ increasing from bottom to top.
Solid lines represent the open models O3-O5, with $\Omega_0$
increasing from bottom to top; the O1 and O2 models produce
no clusters more massive than $10^{13.5}h^{-1}M_\odot$.
The horizontal solid line marks the space density of median $R=1$
Abell clusters, $4\times 10^{-6} h^3 {\rm Mpc}^{-3}$.
The box indicates the mass of observed clusters with this space density 
[$M(R_A)=(3.0-5.5) \times 10^{14} h^{-1}M_\odot$,
$M(R_A/3)=(1.0-1.8) \times 10^{14} h^{-1}M_\odot$],
including a rather generous error allowance as
discussed in the text.  The model curves that do not pass through the
boxes in both panels represent unacceptable models.  
Open circles show Bahcall \& Cen's 
(1993) analytic fit to their estimate of the cluster mass function.
}
\label{fig:cmf}
\end{figure*}

\section{The cluster mass function}

The space density of massive clusters is a sensitive function of
$\Omega_0$ and $\sigma_8$.  Since these two parameters vary in 
concert for DMR-normalized CDM models, the cluster abundance
can provide a tight constraint on $\Omega_0$.  More important still,
the abundance of clusters as a function of mass can be measured
directly from N-body simulations in a way that is insensitive
to uncertain assumptions about galaxy formation.  We 
must assume that galaxies form efficiently enough in cluster-mass
potential wells that these are indeed identified as galaxy clusters,
but this seems more plausible than the alternative assumption that
the universe contains many cluster-mass dark halos that harbour few
galaxies and little X-ray emitting gas.
Most of the uncertainty in using the cluster abundance as a 
cosmological test comes from the systematic uncertainties in
observational estimates of cluster masses, a point to which we will
return later in this Section.

The predicted present-day abundance of clusters in our \dmr
models is given in Fig.~\ref{fig:cmf}. The top panel shows the abundance of
clusters as a function of the mass contained within the Abell radius,
$\ra=1.5 \hmpc$, of each cluster. Following White et al. (\shortcite{wef93}) 
we also plot, in the lower panel, the abundance as function of
the mass within $\ra/3$, since this can often be more robustly
estimated from observations (e.g., van Haarlem, Frenk \&\ White 1997). 
In both cases cluster centres were identified
using a friends-of-friends group finder with a small value of the
linking length, $0.17/\bar n^{1/3}=306 \hkpc$.  
This value picks out the high density knots
of the mass distribution with overdensity of approximately
$1000$. Masses were then measured within spheres centered on the
centre-of-mass of these knots.  Double counting was eliminated by
ranking the cluster centres by the mass of the friends-of-friends
group and assigning particles to the most massive within $\ra$. This
is a small correction, and the results are very similar if one simply excludes
clusters that lie within $\ra$ of a more massive cluster.

In all models, the abundance of clusters declines rapidly with increasing
mass. Clusters are, on average, more massive in models with larger
$\Omega_0$. The number density of clusters of a given mass increases
rapidly with $\Omega_0$ (especially for low $\Omega_0$), and many more
clusters are formed in flat universes than in open universes with the same
value of $\Omega_0$. For example, there are $\sim 6$ times fewer clusters
with $M(\ra)\geq 3\times 10^{14}\hMo$ in the flat $\Omega_0=0.4$ model and
$\sim 25$ times fewer in the open $\Omega_0=0.4$ model than there are in
the $\Omega_0=1$ E1 model.  Flat models with $\Omega_0 \lsim 0.1$ and
open models with $\Omega_0 \lsim 0.3$ produce a negligible abundance of
rich clusters. 

The cluster abundance has been applied as a cosmological test by,
among others, White et al. (\shortcite{wef93}), Viana \& Liddle 
(\shortcite{vl96}), Eke \etal (\shortcite{ecf96}), 
Oukbir \& Blanchard (\shortcite{ob96}),
Oukbir, Bartlett, \& Blanchard (\shortcite{oukbir96}),
Kitayama \& Suto 
(\shortcite{kitayama96}), and G\'orski et al. (\shortcite{gorski96b}). 
These authors compared model predictions based on the Press \& Schechter 
(\shortcite{ps74}) formalism (or generalizations thereof) with the observed 
abundance of clusters measured as a function of mass in the first study 
and as a function of X-ray temperature in the later studies.\footnote{
The first four sets of authors expressed the requirement that a model 
should reproduce the observed cluster abundance in terms of an acceptable 
range of values for $\sigma_8$ as a function of $\Omega_0$.  Comparisons of  
these with the values of $\sigma_8$ required by the DMR normalization are 
given, for example, in figure~13 of Eke \etal (\shortcite{ecf96}),
figure~4 of Viana \& Liddle (\shortcite{vl96}) and 
tables 9-12 of G\'orski et al. (\shortcite{gorski96b}). 
On the other hand, Kitayama \& Suto (\shortcite{kitayama96}) computed cluster
temperature and luminosity functions for \dmr models and compared these to 
the observations.}
The advantage of using a grid of N-body simulations over the 
more flexible analytic approach is that cluster masses can be directly
measured in the simulations, thus bypassing the need to assume a model
relating predicted cluster masses to either measured masses or X-ray
temperatures.

To compare the predictions in Fig.~\ref{fig:cmf} to observations
requires a measurement of the abundance of clusters in a specified
mass range. The open circles in the figure show Bahcall and Cen's
(\shortcite{bc93}) analytic fit to their
estimated mass function of groups and
clusters in the local Universe.  In agreement with their conclusions,
we find that a flat model with $\Omega_0\simeq 0.3$ provides a good
match to their data, whereas a \dmr Einstein-de-Sitter model with
fiducial parameter values (Model E1) overpredicts the abundance of clusters
by a large factor. These conclusions, however, are uncertain because
of the difficulties inherent in defining a complete sample of clusters
over a large range of masses. Further systematic uncertainties could
be introduced by the heterogenous mass estimators employed by Bahcall
\& Cen. [We note, however, that, given the uncertainties, Bahcall \& Cen's 
mass function is in fairly good agreement with the temperature function 
derived by Eke \etal
(\shortcite{ecf96}) from Henry \& Arnaud's (\shortcite{ha91}) X-ray data
and a simple model relating cluster masses and X-ray temperatures.] A
more reliable comparison between models and observational data may be
made by considering a single robust statistic characterising the cluster 
abundance, rather than the mass function as a whole. 
The {\it median} mass of Abell clusters with richness class $R\ge 1$,
advocated by White \etal  (\shortcite{wef93}), has been shown to be 
robust to the distortion of the shape of the measured cluster mass function
caused by line-of-sight projections (van Haarlem \etal 1997).
The abundance of $R \geq 1$ clusters, 
$N_{\rm A}=8\times 10^{-6}h^3 {\rm Mpc^{-3}}$, is well established from X-ray 
and optical studies (Bahcall \& Soneira \shortcite{bs83}; 
Scaramella et al. \shortcite{scaramella91};
Efstathiou et al. \shortcite{edsm92}). 
The results of the comparison do not depend sensitively on the adopted value 
of $N_A$ because the predicted number density of clusters declines very steeply
with mass.  For the same reason, however, the results are quite sensitive to 
the adopted value of the median mass.

White \etal (\shortcite{wef93}) give the range 
$M_{\rm clus}(\ra)=(4.2-5.5)\times10^{14} h^{-1}\Mo$ for the median mass
within the Abell radius of $R\ge 1$ Abell clusters.  The lower limit
comes from estimates using X-ray data and the upper limit from
estimates using velocity dispersions of cluster galaxies. As White
\etal emphasize, the statistical significance of this range is
difficult to quantify because most of the uncertainties arise from
systematic errors in the mass determinations. The quoted range is
meant to take at least some of these systematics into account.

We adopt a slight modification of this range. We retain the 
upper mass limit advocated by White \etal (\shortcite{wef93}),
but reduce the lower mass limit to $3.0\times10^{14} h^{-1}\Mo$.
This lower value follows from two considerations. First, the reanalysis
by Eke \etal (\shortcite{ecf96}) of the X-ray data compiled by Henry \& Arnaud 
(\shortcite{ha91}) yields an X-ray temperature 
$3.3\;$keV for a median $R=1$ cluster
rather than the $3.6\;$keV
adopted by White et al.\ (\shortcite{wef93}). 
Second, White \etal extrapolated the cluster
mass from the virial radius to $\ra$ assuming an $r^{-2}$ density profile.
This extrapolation amounts to a factor of $1.35$ in mass. If cluster density
profiles are significantly steeper at the virial radius, as suggested
by Navarro, Frenk \& White (\shortcite{nfw95}),
then a more modest factor is appropriate. Hence we adopt 
$3.0\times10^{14} h^{-1}\Mo$ as a conservative lower limit.
For masses within $R_A/3$, we simply reduce the mass range by a factor
of three at the upper and lower ends, which is correct if 
cluster density profiles are $\propto r^{-2}$.  In practice, the
estimated masses within $R_A/3$ are probably more robust than those
within $R_A$, since the X-ray data typically do not extend all the
way to $1.5h^{-1}\;$Mpc.
However, we reach similar conclusions about the viability of models
using either measure.

In addition to the traditional methods for estimating cluster masses,
based on the kinematics of their galaxies and/or on the hydrodynamics
of their X-ray emitting gas, a relatively new method, based on the
gravitational lensing properties of the cluster dark matter, has
recently been applied to several clusters. The extent to which all
three methods give consistent results is still a matter of debate. All of them
require a number of assumptions of varying degrees of
plausibility (see, e.g., White \etal \shortcite{wef93}). With standard
assumptions, estimates based on velocity dispersions often give
somewhat larger masses than estimates based on X-ray data.
There are examples for which the
mass inferred from gravitational lensing differs significantly from the mass
inferred from X-ray or dynamical data (e.g., \cite{fahlman94})
and others where the agreement is good (e.g., \cite{squires96}). At present, 
the sample of clusters analyzed by gravitational lensing is small, and it 
consists predominantly of clusters at  intermediate redshift, where the 
traditional methods are particularly uncertain. 
An important argument for the robustness of dynamical estimates
based on galaxy motions is the fact that blue and red galaxy populations
yield compatible masses, despite having different density and 
velocity dispersion profiles (\cite{carlberg96}).

From Fig. 4, we see that the \dmr CDM models that
produce an acceptable cluster abundance are the
open O4 and~O5 models, the flat L3 and~L4 models, and, by construction,
the tilted $\Omega_0=1$ model E2. Thus we estimate that $\Omega_0 \gap 0.25$ 
is required to produce the observed abundance of rich clusters in the 
spatially-flat models and that this lower limit rises to
$\Omega_0 \gap 0.4$ for open models. The corresponding upper limits are
$\Omega_0 \lap 0.4$ for the flat models and $\Omega_0 \lap 0.5$
for open models. The $\Omega_0=1$ model with the scale-invariant $n=1$ 
primordial power spectrum, E1, produces an abundance of $R\geq 1$ clusters
that is $20$ times too high.

These constraints are in accord with more approximate ones derived earlier
(e.g., \cite{ecf96,kitayama96,gorski96b}). Note, however, that our analysis
does not account for the known uncertainties in estimates of $t_0$,
$\OmegaB$, and the DMR normalization. These uncertainties will somewhat 
broaden the above allowable ranges of $\Omega_0$. For example,
G\'orski et al.  (\shortcite{gorski96b}) compare \dmr values of 
$\sigma_8$ to those 
deemed necessary by Eke et al. (\shortcite{ecf96}) to explain the observed 
cluster temperature function, and accounting for the 2-$\sigma$ uncertainty 
in the DMR normalization and some of the uncertainty in $t_0$ and $\OmegaB$,
they find an allowable range of
$0.3 < \Omega_0 \lap 0.6$ ($\sim$2-$\sigma$) for the open-bubble 
inflation model.

\begin{figure*}
\centering
\centerline{\epsfxsize = 5.0 truein \epsfbox[5 205 550 775]{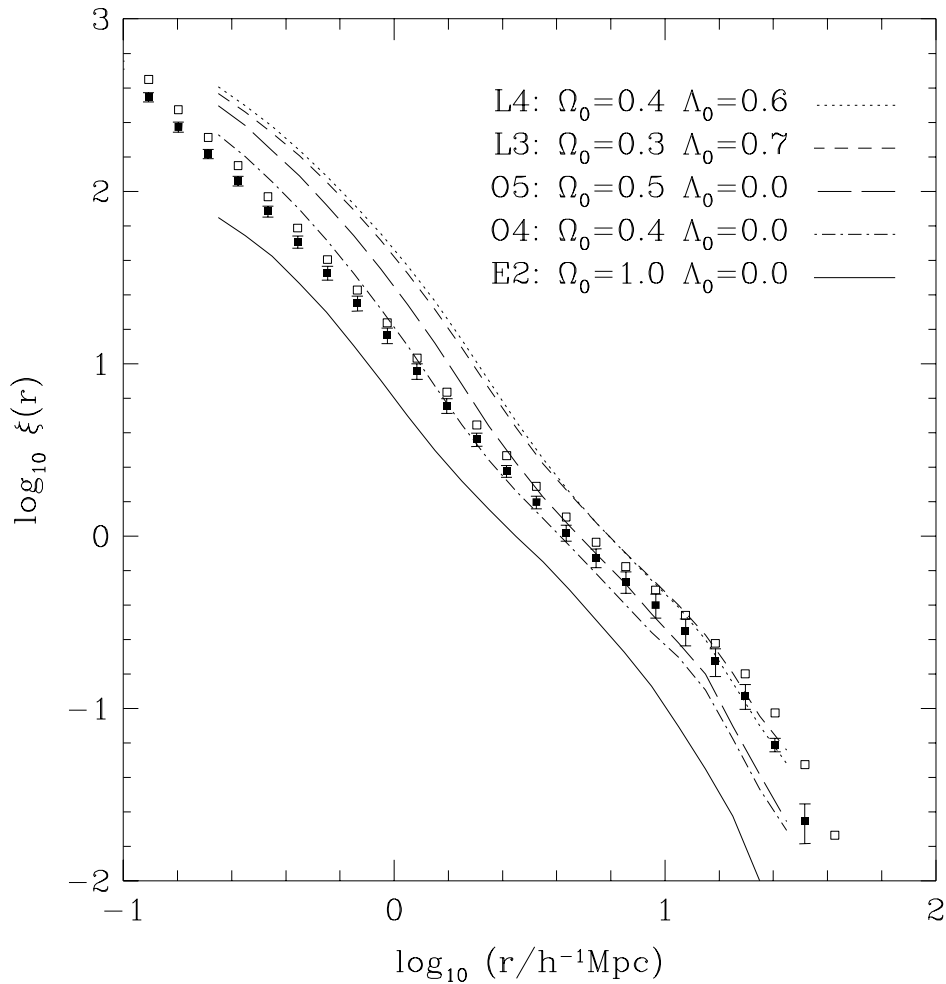}}
\caption{Mass autocorrelation functions for the five models from our
suite of N-body simulations that yield acceptable cluster masses.  
Short-dashed and dotted
lines show the flat models with $\Omega_0=0.4$ and 0.3, respectively,
long-dashed and dot-dashed lines show the open models with $\Omega_0=0.5$
and 0.4, respectively, and the solid line shows the tilted
$\Omega_0=1$ model.  Results are shown from $r=35\hmpc$, 1/10 of the
box size, down to $r=0.2\hmpc$, about twice the effective gravitational
softening length $\epsilon=0.09\hmpc$. Solid squares with error bars show 
the real space galaxy correlation function inferred by Baugh (1996) 
from the APM angular correlation data assuming that it remains fixed in 
comoving coordinates. Open squares show the galaxy correlation function 
derived assuming linear theory evolution of the correlation function with 
redshift --- statistical errors on these points (not shown)
are similar to those on the solid points. 
}
\label{fig:xi}
\end{figure*}

\begin{figure*}
\centering
\centerline{\epsfxsize = 5.0 truein \epsfbox[5 205 550 775]{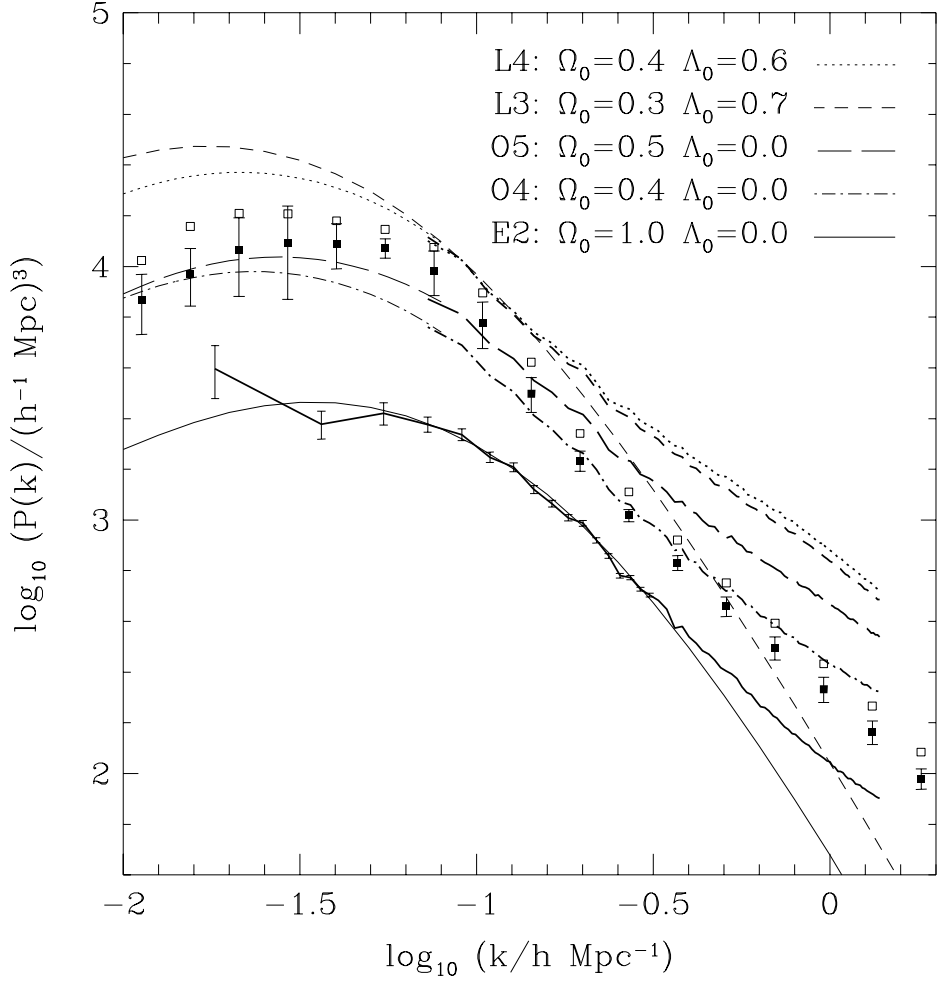}}
\caption{Mass power spectra for the five models shown in
Fig.~\ref{fig:xi}, with the same correspondence between model and line type.  
For each model the heavy curve shows the results of the N-body simulation
and the light curve the corresponding linear theory power spectrum. 
For the model E2, we indicate the 1-$\sigma$ errors due to the finite number
of modes in the simulation cube.
The smallest $k$ value in the E2 N-body $P(k)$ is the fundamental frequency
of the $345.6\hmpc$ simulation box; the largest is $0.8 k_N$, where
$k_N$ is the Nyquist frequency of our $192^3$ FFT grid. 
For the other models we omit the error bars and do not plot the first
three, noisy data points. For models E2 and L3 the linear theory curves
are plotted over the full range of $k$ to illustrate the effect of 
non-linearities at small scales. For the other models the linear theory
power spectra are just plotted as an accurate extrapolation of the model
power spectra to small $k$.
The filled squares with 
error bars show the real space galaxy power spectrum inferred by Baugh 
\& Efstathiou (1993) from the APM angular power spectrum, assuming
that clustering remains fixed in comoving coordinates. Open squares 
(without error bars) show the corresponding galaxy power spectrum assuming 
linear theory evolution of clustering with redshift 
for an $\Omega_0=1$ universe. 
}
\label{fig:psp}
\end{figure*}

\begin{figure*}
\centering
\centerline{\epsfxsize = 5.0 truein \epsfbox[5 205 550 775]{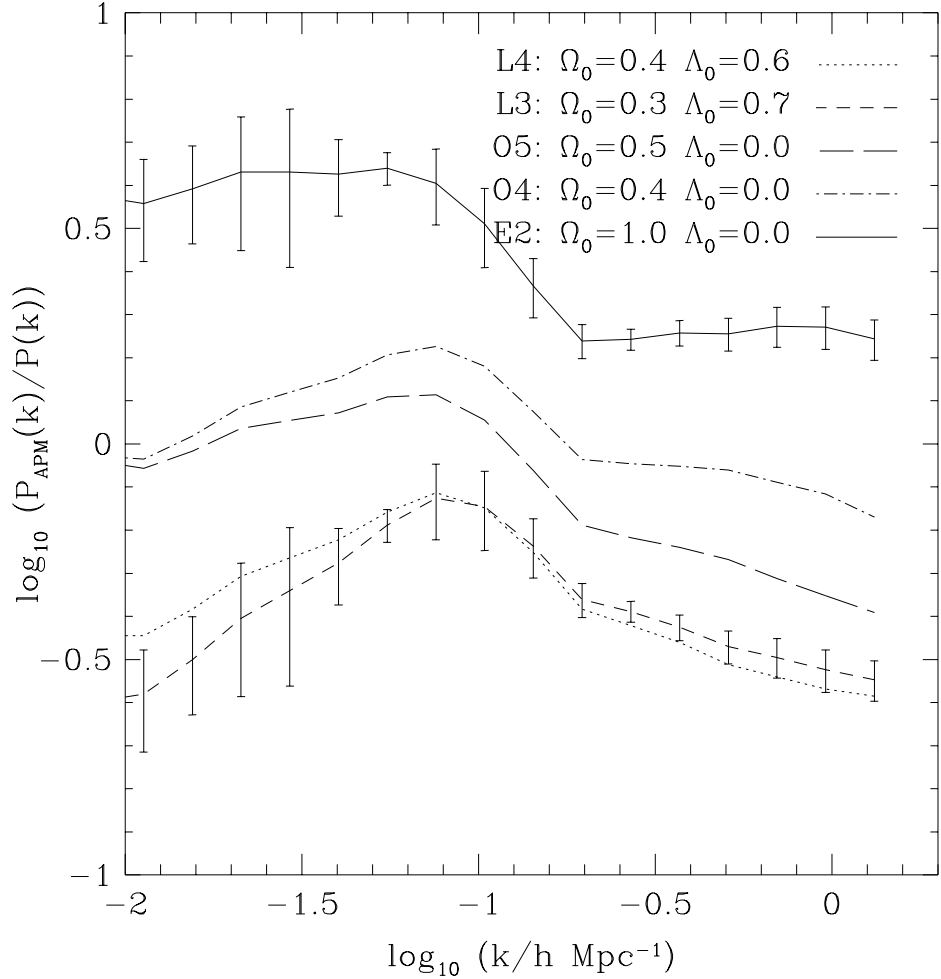}}
\caption{ The ratio of the APM galaxy power spectrum, as inferred by
Baugh \& Efstathiou (1993) for clustering fixed in comoving coordinates,
to the mass power spectrum of each of the five models shown in
Fig.~\ref{fig:xi} and~\ref{fig:psp}. The correspondence between model
and line type is the same as in the previous two figures. The error bars 
(for clarity only plotted on two of the models) indicate only the statistical
error arising from the APM galaxy power spectrum. 
Uncertainties associated with the assumed evolution of clustering and APM 
selection function (see text) can lead to a systematic shift of $\sim 0.2$ in
$\log(P_{APM}(k)/P(k))$. Also, for $k \lsim 0.06 h {\rm Mpc}^{-1}$,
systematic errors in the APM angular correlation function probably result 
in errors in $\log(P_{APM}(k)/P(k))$
which are larger than the plotted statistical errors
(Maddox et al. 1996).
}
\label{fig:pspr}
\end{figure*}

\begin{figure*}
\centering
\centerline{\epsfxsize = 5.0 truein \epsfbox[35 175 515 640]{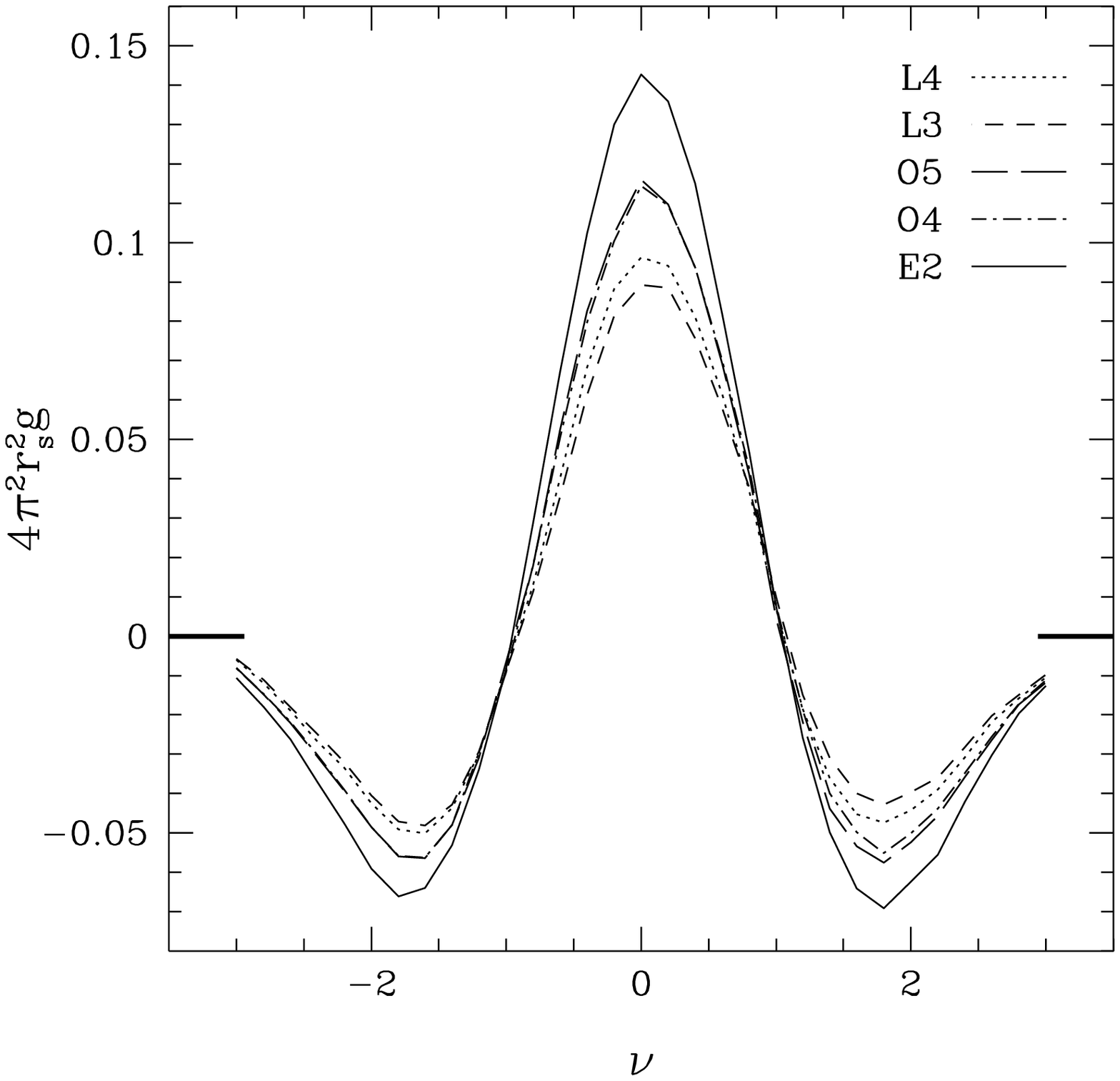}}
\caption{Topology of isodensity contour surfaces
for the five models shown in Fig.~\ref{fig:xi}.
The topology is measured from
mass density fields that have been smoothed with a Gaussian filter of radius
$r_s=4.5\hmpc$. The scaled genus-per-unit-volume, $4\pi^2 r_s^3 g$,
is plotted against the value of $\nu$ defined in eq.~(\ref{eqn:nudef}).
}
\label{fig:genus}
\end{figure*}

\section{Other measures of clustering}

We now compute some statistical characteristics of the real space mass
distribution for the five models from our suite that produce
observationally acceptable cluster abundances.  The statistics of the
galaxy distributions in these models will depend on the relation
between galaxies and mass (``bias'' or ``antibias''), and, if they are
measured in redshift space, on the distortions caused by peculiar
motions. The measures presented here will give some sense of how the
models compare to each other and indicate what kinds of bias would be
needed to obtain acceptable fits of our N-body results to observational data.

Figs.~\ref{fig:xi} and~\ref{fig:psp} show the mass autocorrelation
functions $\xi(r)$ and power spectra $P(k)$ from these five models.
We also show estimates of the real space galaxy autocorrelation
function and power spectrum derived by Baugh (\shortcite{baugh96}) and
Baugh \& Efstathiou (\shortcite{baugh93}) from the angular correlation
function of the APM galaxy catalogue (\cite{apm90}).  Since the APM
catalogue is fairly deep, the inferred $\xi(r)$ and $P(k)$ depend
somewhat on the assumed redshift evolution of clustering.  We show
Baugh's and Baugh and Efstathiou's results for two different
assumptions, linear theory evolution for $\Omega_0=1$ (open squares)
and clustering fixed in comoving coordinates (filled squares).  These
assumptions probably bracket the true evolution, and the difference
between them illustrates the systematic uncertainty associated with
this effect.  The values $\sigmag$, the rms fluctuation of the density of
galaxies in spheres of radius $8 \hmpc$,
implied by these two results
are $\sigmag \approx 0.82$ (fixed in comoving coordinates) and
$\sigmag \approx 0.92$ (linear evolution).  The inferred clustering
amplitude, which we can characterize by the value of $\sigmag$, is
also subject to other systematic uncertainties. Uncertainties in the
APM selection function and in the assumed cosmological model both
introduce uncertainties at the level of a few percent. Furthermore, 
systematic errors exist in APM angular correlation function at some
level.  Maddox, Efstathiou \& Sutherland (\shortcite{apmIII}) analyse a wide range of
possible systematic errors and apply corrections for recognized
problems with star-galaxy classification at faint magnitudes and other
small systematic errors. These corrections lead to $\sigmag \approx
0.96$ for the same cosmological model and redshift distribution
assumed above with a clustering evolution intermediate to the two
cases above.  The error bars shown on the APM data in
Figs.~\ref{fig:xi} and~\ref{fig:psp} indicate the statistical errors
and are estimated from the dispersion among four nearly equal zones of
the APM survey.  For the most part, the sources of systematic error
discussed above tend to alter the amplitude of $\xi(r)$ and $P(k)$
without affecting their shape. However, the points at largest
separation ($r \gsim 40 \hmpc$) and lowest wavenumber ($k \lsim 0.06{h
{\rm Mpc}^{-1}}$) probably have uncertainties that are dominated by 
systematic errors (Maddox et al. \shortcite{apmIII}).

Fig.~\ref{fig:xi} shows the mass autocorrelation functions $\xi(r)$
from the five acceptable models, measured from twice the effective
gravitational softening length, $2\epsilon=180\hkpc$, out to
$35\hmpc$, $\sim 1/10$ of the simulation box size.  Fig.~\ref{fig:psp}
shows the mass power spectra of the five models.  On the largest
scales the linear theory power spectra are plotted and these are then
continued to higher $k$ using estimates from the N-body
simulations. We measure the model power spectra in the N-body
distributions by FFT after cloud-in-cell (CIC) assignment of the
particle distribution onto a $192^3$ grid.  We correct the FFT power
spectrum for the effect of CIC convolution.  In two cases (models E2
and L3), the linear theory power spectrum is shown for all $k$ so that
the effect of non-linear evolution on $P(k)$ can be seen. 
Error bars are shown for model E2, which illustrate the uncertainty in
estimating $P(k)$ on very large scales where there are few modes in the
simulated cube.  The relative biases required in the models if they
are to reproduce the observed galaxy clustering are shown most clearly
in Fig.~\ref{fig:pspr}, which shows the ratio of APM galaxy power
spectrum to the model mass power spectra.

%

The L4 and L3 models, which have $(\Gamma,\sigma_8)=(0.21,1.1)$
and $(0.17,1.05)$, respectively, have mass correlation functions 
which trace the APM galaxy correlations fairly well at large separations, 
$r \gap 10\hmpc$.  However, as previously noted (e.g., 
\cite{apm_nature,klypin96,apmIII,peacock96}), their mass correlation
functions steepen considerably at $r \lap 4\hmpc$, rising above
the APM galaxy data by a factor $\sim 3$ at $r=1\hmpc$.  
To match the observed galaxy correlation function, these models would
require a strong antibias between galaxies and mass on these scales
but could not tolerate much antibias or bias at large scales.
The O5 model has a higher $\Gamma$ (0.27) and a lower $\sigma_8$ (0.9).
Its mass correlation function traces the APM galaxy data for scales
where $\xi(r) \sim 1$.  The mass correlation function again steepens
at smaller scales and rises above the APM galaxy data, though not as much
as for the $\Lambda$ models.  It falls below the APM data at
scales larger than $r \sim 10\hmpc$.  To fit the data, this model
would require a positive bias at large scales and an anti-bias at
small scales.  
The O4 model, with $(\Gamma,\sigma_8)=(0.23,0.75)$,
traces the APM galaxy data fairly well for $r\lap 5\hmpc$.  Because mass
fluctuations are less nonlinear in this model, the mass correlation
function does not show the marked steepening inside the correlation
length seen in the earlier models.  However, the correlation 
function falls below the APM data at large separations, so the 
O4 model would require positive bias on these scales in order to match 
the data.  
The E2 model, with $(\Gamma,\sigma_8)=(0.45,0.55)$ and $n=0.803$,
requires a positive bias on all scales to match the APM galaxy data, as
expected.  Again the low amplitude means that there is no marked
steepening of the slope inside the correlation length.  The
required bias is roughly independent of scale except at the largest
separations, where the E2 correlation function curls away from
the data. 
The APM data are well fit on these scales by a model
with an $n=1$ primordial power spectrum and $\Gamma \sim 0.15-0.2$.
The E2 model fits better than an $n=1$ model with the same value
of $\Gamma=0.45$, but it nonetheless
yields smaller correlations than the APM data on scales of $30-100\hmpc$.

Overall, the power spectrum comparison confirms
our conclusions from the $\xi(r)$ comparison (as expected, since the
two measures are a Fourier transform pair).  The E2 model requires
a positive bias on all scales to match the APM galaxy data; the O4 and O5
models require a positive bias on large scales
($k \sim 0.05\;h\;$Mpc$^{-1}$), with O5 also requiring antibias
on small scales; and the L3 and L4 models require a weak antibias
on large scales (previously noted in linear theory mass $P(k)$ and CMB 
anisotropy computations, e.g.,
\cite{stompor95,os95,ratra96,ganga96c}) and a strong antibias on small scales.

The correlation function and the power spectrum quantify the rms
amplitude of fluctuations as a function of scale, but they 
contain no further information about the global structure of the 
density field.  As a complementary statistic, we consider the 
topology of isodensity contour surfaces, following the methodology
of Gott, Melott, \& Dickinson (\shortcite{gmd86}) and
Gott, Weinberg, \& Melott (\shortcite{gwm87}).
We smooth the mass density field by convolution with a Gaussian 
filter of radius $r_s=4.5\hmpc$ (corresponding to
$\lambda=4.5\sqrt{2}=6.4\hmpc$ with the smoothing filter definition
of \cite{gwm87}), then measure the genus $G_{\rm s}$ of isodensity contours
at a range of threshold densities.  Fig.~\ref{fig:genus} plots
the genus per unit volume, $g=G_{\rm s}/(345.6\hmpc)^3$,
multiplied by the scaling factor $4\pi^2 r_s^3$, as a function of
$\nu$, which is defined implicitly in terms of the contour's 
enclosed fractional volume $f$ through the equation
\begin{equation}
f = \frac{1}{\sqrt{2 \pi}}\,\int_\nu^\infty e^{-x^2/2} dx.
\label{eqn:nudef}
\end{equation}
An advantage of the volume-weighted procedure implied
by equation~(\ref{eqn:nudef}) is that it makes the ``genus curve''
insensitive to biasing or antibiasing of the galaxy distribution,
in contrast to $\xi(r)$ or $P(k)$.  To the extent that biasing
preserves a monotonic relation between the smoothed mass density
and smoothed galaxy density, it does not alter the genus curve, 
even if the relation is nonlinear.  In practice, plausible biasing
models can slightly distort the genus curve (\cite{pg91}),
e.g., by attenuating the walls of galaxies that separate underdense
regions, but we expect that the genus curves of galaxy density
fields in these models would be close to the genus curves of the
mass density fields shown in Fig.~\ref{fig:genus}.

A Gaussian random field with power spectrum $P(k) \propto k^n$ has
a mean genus curve
\begin{equation}
4\pi^2 r_s^3 g = \left(\frac{3+n}{6}\right)^{3/2} (1-\nu^2)e^{-\nu^2/2}
\label{eqn:gcurve}
\end{equation}
(\cite{dor70}; \cite{adler81}; \cite{bbks}; \cite{hgw86}).
With the volume-weighted convention for labelling contours,
mildly nonlinear gravitational evolution has only a small effect
on the shape of the genus curve, though it does tend to lower the
overall amplitude (\cite{mwg88}; \cite{pg91}).
The symmetric, W-shaped form of the curves in Fig.~\ref{fig:genus}
reflects the Gaussian nature of the initial fluctuations in all
of these models.  The different amplitudes reflect the differences
in the logarithmic slope of the primordial power spectrum on the
$4.5\hmpc$ smoothing scale (eq.~[\ref{eqn:gcurve}]) and the 
differing degrees of nonlinear dynamical evolution corresponding
to different values of $\sigma_8$.

A disadvantage of the topology statistic is that measurements from
``small'' redshift samples (e.g., flux-limited redshift surveys of
a few thousand galaxies) suffer significant systematic biases
because of the finite survey volume and the unknown boundary
conditions for defining the smoothed density field
(see, e.g., the discussion by \cite{pw97}).
These can be taken into account by comparing observational results directly
to those of artificial catalogues drawn from N-body simulations,
but this level of detailed modelling is beyond the scope of this paper.
We have not included observational data in Fig.~\ref{fig:genus}
because there is no simple way to incorporate the effect of these
sample-dependent systematic biases, but studies comparing
mock catalogues to a number of galaxy redshift surveys show that
the observed topology is generally consistent with models that
have Gaussian initial fluctuations with a $\Gamma \approx 0.25$
power spectrum (\cite{gott89,park92,moore92,vogeley94,pw97}).
Current topology data are probably not sufficient to
distinguish between the models illustrated in Fig.~\ref{fig:genus},
but future redshift surveys like the 2dF and SDSS
(\cite{colless96}; \cite{gw95}) should yield precise topology 
measurements with minimal systematic errors that can easily
distinguish most of these models from each other (the exception
being O4 and O5, which have nearly identical genus curves).
Comparisons with high-precision data will also need to incorporate
the impact of galaxy biasing, but we expect this to be relatively
small for the reasons discussed earlier.

\section{Discussion}

We have used large N-body simulations of CDM models to constrain the range
of model parameters which, given some simple assumptions, are
consistent with the cosmic microwave background anisotropies measured
by \cdmr and the abundance of rich galaxy clusters in the local universe. We
have then investigated in detail some basic clustering statistics for the
dark matter in a selection of acceptable models.

The combination of DMR anisotropy and cluster abundance provides an
appealing way to constrain the parameters of a CDM universe because, by
avoiding explicit reference to galaxies, it bypasses the highly uncertain
connection between the large-scale distributions of galaxies and mass.
Nevertheless, this test does have a number of important limitations. The
extrapolation of the fluctuation amplitude from the DMR scale to the
cluster scale, $r\approx 8 \hmpc$, requires an assumption about the
shape of the primordial mass power spectrum emerging from inflation,
e.g., a scale-invariant ($n=1$) spectrum in flat models or the simplest
open-bubble inflation spectrum (which is close to $n=1$ on scales much
smaller than the present Hubble length).
This shape is not tightly constrained by the DMR data, and 
because of the large lever arm between the
DMR scale and $8 \hmpc$, small changes in the spectral shape can
produce large changes in
the inferred value of $\sigma_8$. This uncertainty is compounded by the
fact that in standard flat inflation models even small deviations from 
$n=1$ can result in a significant contamination of the measured anisotropy 
signal by tensor modes, which produce gravity waves rather than density
fluctuations (e.g., \cite{salopek92,crittenden93}). 

The determination of $\sigma_8$ from the observed cluster abundance has the
great advantage of being essentially independent of the shape of the power
spectrum if $\Omega_0=1$ and of depending only very weakly on it if
$\Omega_0<1$.  The dependence on $\Lambda_0$ is also weak; for $\Omega_0 >
0.2$, the difference in $\sigma_8$ between flat and open models with the
same value of $\Omega_0$ is always less than 10 per cent (\cite{ecf96}).
The main uncertainty stems from the difficulty in estimating cluster masses
reliably. The expected number of clusters declines very steeply with
cluster mass and, as a result, the estimate of $\sigma_8$, while not
requiring a particularly accurate measurement of the abundance of clusters,
is very sensitive to the estimates of cluster mass.  In this paper, we have
adopted a slightly wider range of cluster masses than than quoted by
White \etal (\shortcite{wef93}) 
for the median mass of $R\ge 1$ Abell clusters within the
Abell radius $\ra =1.5 \hmpc$: $M_{\rm clus}(\ra)=(3.0-5.5) h^{-1}\Mo$. 
The lower
end of this range corresponds to our revised estimate from X-ray data and
the upper end to the estimate from dynamical data.

In our series of low-$\Omega_0$ CDM simulations, we assume the simplest
primordial inflation power spectra, a Hubble constant that yields
$t_0 \approx 14$ Gyr (flat models) or 12 Gyr (open models), a baryon
density parameter $\OmegaB =0.0125h^{-2}$, and no influence of gravity waves
or mild early reionization on the DMR anisotropies.
With these assumptions, there is only a rather narrow range of DMR-normalized
CDM models that produce the observed abundance of rich galaxy
clusters. Flat models ($\Omega_0 + \Lambda_0 =1$) require
$\Omega_0=0.25-0.4$, whereas open models require $\Omega_0=0.4-0.5$.  Most
plausible departures from our assumptions tend to drive these estimates to
higher values of $\Omega_0$. 
For example, lowering $H_0$ in order to increase $t_0$
reduces the value of $\sigma_8$ implied by CMB 
anisotropies but has little effect on the estimate of $\sigma_8$ from the
cluster abundance.  Thus a lower $H_0$ requires a higher $\Omega_0$, and
$\Omega_0=1$ is acceptable if $h\approx 0.25$. Increasing the baryon
fraction slightly above our standard value has an effect in the same
direction, but the impact is substantially weaker than the impact
of lowering $H_0$ unless $\OmegaB$ becomes a large fraction of
$\Omega_0$.  Lowering the baryon fraction would go in the direction
of allowing lower $\Omega_0$, but here the effect is quite small
because our acceptable models already have a low ratio of $\OmegaB/\Omega_0$.
Our adopted DMR normalizations, based on the \cdmr two-year data,
tend to be higher (by roughly 10\% or +1-$\sigma$) than the central
values estimated from the 4-year data (see discussion in Section 2.1).
Lowering the normalizations
would reduce the value of $\sigma_8$ for a given $\Omega_0$, so again
a higher $\Omega_0$ would be required to match the observed cluster abundance.

Generalizations of the simplest inflation models most easily
lead to redder primordial power spectra, e.g., to an effective
index $n<1$ in flat models.  
Lowering $n$ reduces the $\sigma_8$ inferred from DMR-normalization
both because it changes the extrapolation of fluctuation amplitudes
to cluster scales and because, at least in power-law inflation
models, such tilted spectra are accompanied by the production of gravity
waves.  Any change towards a redder inflation power spectrum 
thus raises the lower limit on $\Omega_0$.
Reducing $n$ to about 0.8 is a simple way to make
$\Omega_0=1$ compatible with the DMR data and the observed cluster
abundance, though any $\Omega_0=1$ model remains difficult to reconcile
with the high baryon fraction in rich clusters and with
current observational estimates of $t_0$ and $H_0$.

While our adopted range of median cluster
masses is fairly conservative, we cannot exclude the possibility that this
estimate may be systematically too low or too high. The gravitational
lensing data suggest that, if anything, the former is more likely.  In this
case the observed cluster abundance would require a higher value of
$\sigma_8$ and hence a higher value of $\Omega_0$ for consistency with the
DMR data.

We conclude that, under the assumptions discussed above,
$\Omega_0 \gsim 0.25$ is required to produce enough massive galaxy 
clusters in spatially-flat, DMR-normalized CDM models. In open models, the
value of $\sigma_8$ implied by the DMR data drops so fast with decreasing
$\Omega_0$ that the lower limit rises to $\Omega_0 \gsim 0.4$. The statistical
significance of these limits is difficult to quantify because the errors
are dominated by the systematic uncertainties just discussed. 
Our conclusions agree well with previous N-body work 
(\cite{frenk90,bc93,wef93,ecf96}) and with
semianalytic results based on the Press-Schechter model 
(\cite{wef93,ecf96,kitayama96,vl96}).  However, we are able to set a
more stringent lower limit on $\Omega_0$ because the use of
large N-body simulations removes any uncertainty regarding the relation
between predicted and observed cluster masses.  Related constraints on
cosmological parameters follow from considering the {\it evolution} of the
abundance of clusters (\cite{ecf96,ob96}). Current cluster 
evolution data tentatively indicate $\Omega_0 \approx 0.4-0.5$,
$\sigma_8 \approx 0.75$  (Carlberg et al. 1997a; 
Henry 1997 private communication), consistent with the constraints 
derived here for open models.

There are possible changes to our assumptions or parameter choices
that would weaken these lower bounds on $\Omega_0$.
These include raising $H_0$ (and consequently lowering $t_0$)
or going to the high end of the error range for the DMR-normalization.
Taking these and other uncertainties into account,
G\'orski et al.\ (\shortcite{gorski96b}) conclude that open-bubble
inflation models with $\Omega_0$ as low as 0.3 
could be consistent with current observations.
Lower $\Omega_0$ is also allowed if the inflation model produces
a bluer, ``anti-tilted'' primordial power spectrum, though in some cases
the effect of such a change on the DMR-normalized $\sigma_8$ would
be partly cancelled by the contribution of gravity waves to the
DMR anisotropies.  Open CDM models with $n>1$ 
(on scales well below the Hubble length) have been advocated
by Bunn \& White (\shortcite{bw96}) and White \& Silk (\shortcite{ws96}).
In these studies the anti-tilted spectrum is imposed {\it ad hoc};
it has not yet been shown that such a spectrum can be produced by
quantum fluctuations in an open-universe inflation model.\footnote{
A recent study (Garc\'{\i}a-Bellido \& Linde 1997) 
indicates that quantum fluctuations in an open-bubble
inflation model with three scalar fields might produce an anti-tilted
generalization of the usual single-scalar-field open-bubble inflation
model primordial power spectrum (Ratra \& Peebles 1994,1995; Bucher 
et al. 1995; Yamamoto et al. 1995).}

Matching the \cdmr anisotropy amplitude and the abundance of rich
clusters does not, by itself, guarantee a successful cosmological model. Of
the many criteria available for discriminating amongst cosmological models,
the observed pattern of galaxy clustering remains one of the most widely
used. However, whereas the CMB and cluster abundance tests apply directly
to the distribution of dark matter, the galaxy clustering test requires a
model for the galaxy distribution. Unfortunately, our understanding of
galaxy formation is still so primitive that only highly idealized
mathematical models such as the ``high peak'' model of biased galaxy
formation (\cite{defw85,bbks}) 
have been explored in detail. Cosmological gasdynamic
simulations do not yet have sufficient resolution to follow galaxy
formation except in relatively small volumes
(\cite{co93,katz92,summers95,frenk96}).  Our dark
matter simulations do not address the issue of galaxy formation, but they
do at least provide an accurate description of the statistical properties
of the dark matter distribution on scales $\sim 0.2-35 \hmpc$. Comparing
this mass distribution with the {\it observed} galaxy distribution reveals
the kind of biases that must be present in the galaxy distribution for a
particular cosmological model to be acceptable. 

We find that all DMR-normalized CDM models that successfully
reproduce the cluster abundance require that the galaxy distribution be
biased relative to the mass in a non-trivial fashion. Except in the
$\Omega_0=0.4$, $\Lambda_0=0$ case, the acceptable low-$\Omega_0$ models,
with or
without a cosmological constant, produce mass distributions that are {\it
significantly more strongly clustered} than the observed galaxy
distribution on small and intermediate scales. These models thus require
the galaxy distribution to be antibiased relative to the mass on these
scales. On scales $\gsim 10 \hmpc$, the mass correlations in these models
become comparable to the galaxy correlations. In the $\Omega_0=0.4$,
$\Lambda_0=0$ case, the predicted mass correlations are only slightly
stronger than the observed galaxy correlations on small scales
(and, within the uncertainties, could even be weaker), but they fall
well below the observed $\xi(r)$ on scales larger than $\sim 10 \hmpc$. 
This model thus requires (mild) antibias on small scales and positive bias 
on 
large scales. Finally, the tilted $\Omega_0=1$ model requires a strong
positive bias everywhere, and the degree of bias depends on scale.  Beyond 
$\sim 10 \hmpc$, it must rise steeply with scale to compensate for the
rapidly declining large-scale clustering strength. 

Little is known about the physical plausibility of a strong antibias
in the galaxy distribution on small and intermediate scales,
which would be required
for almost all our low-$\Omega_0$ models to match observations. Galaxy mergers
might produce this kind of effect, and preliminary results from a large
programme of cosmological gasdynamic simulations suggest that
galaxies may indeed be antibiased in low $\Omega_0$ models,
although at a lower level than seems required by observations
(\cite{jenkins96}), except possibly for $\Omega_0 \sim 0.4$ open models.
On empirical grounds, however, an antibiased
galaxy distribution in models with $\Omega_0>0.2$ seems difficult
to reconcile with virial analyses of galaxy clusters.  If the
mass-to-light ratio of clusters equals the universal value, then these
imply that $\Omega_0=0.19$, with a 2-$\sigma$ upper limit $\Omega_0<0.33$
(Carlberg, Yee \& Ellingson \shortcite{carlberg97};
Carlberg \etal \shortcite{carlberg96}).
However, if galaxies are over-represented in clusters by a factor $B$, then the
data imply $\Omega_0=0.19B$.  If galaxies were antibiased on
small scales, then we would expect them to be under-represented in clusters, 
implying $B<1$, and therefore $\Omega_0<0.19$ (or $<0.33$ at 2-$\sigma$).

Given these considerations, our results lend support to earlier suggestions 
(\cite{apm_nature,stompor95,os95,ratra96,ganga96c,klypin96,apmIII,peacock96})
that low-density, flat-universe CDM models can more easily fit
the observations if they have a mildly tilted primordial mass power spectrum. 
Flat-$\Lambda$ models with $n \sim 0.9$ could still have reasonable
values of $\Omega_0$ and $t_0$ for a given $h$, and they would require
less small-scale antibiasing in the galaxy distribution.
It is possible but less clear that open CDM models would benefit
from an anti-tilted primordial spectrum, for although this would
allow a lower $\Omega_0$ and higher $t_0$ for a given $h$, it would also
lead to stronger small-scale mass clustering and hence a requirement of
substantial antibias in the galaxy distribution on these scales. A bluer
primordial power spectrum could also bring the small-scale (multipole
$\ell > 300$) CMB anisotropies in open models into conflict with upper
limits (\cite{tucker93}, \cite{ratra96}, \cite{church96},
\cite{ganga96a}), and it might result in a steeper CMB anisotropy spectrum
on intermediate scales ($\ell \sim 200$) than is indicated by current data
(\cite{ganga96c}). Both of these effects, however, could be ameliorated by
mild early reionization.

To illustrate the region of parameter space allowed by current 
observational uncertainties, consider an $\Omega_0 = 0.35$, open-bubble
inflation model with $t_0 = 12\;$Gyr (as advocated by \cite{alcock96}) and
$\OmegaB h^2 = 0.007$ (as advocated by
\cite{songaila94,carswell94,rh96a},b). At the 2-$\sigma$ upper limit, the
\dmr $\sigma_8$ for this model is 0.69 (\cite{gorski96b}), consistent, also
within 2-$\sigma$, with the Eke et al. (\shortcite{ecf96}) cluster
abundance requirement
and with the constraint from cluster mass-to-light ratios that $B > 1.1$ 
(2-$\sigma$, Carlberg \etal 1997b,c). The slightly smaller
$\Omega_0$ compared to the O4 simulation
illustrated in Fig.~5 would
lead to a slightly lower $\Gamma$ and, on large scales, this change would
compensate for the reduction in clustering strength caused by reducing
$\sigma_8$ from 0.75 to 0.69. An open model with these parameters might
thus be consistent with the DMR anisotropies, the evolution of the 
abundance of clusters, the
APM clustering data, and cluster mass-to-light ratios if it had a weak
positive bias on small scales and a somewhat stronger positive bias
on larger scales.  Agreement at this level could probably also be 
achieved by a tilted, flat-$\Lambda$ CDM model (\cite{os95}).

A great variety of observations currently underway or planned for the near
future will significantly sharpen the constraints that we have discussed
in this paper.  Analyses of existing small-angle CMB anisotropy data,
based either on comparisons of observations from many different 
experiments with model predictions for a range of parameter values (e.g., 
\cite{ratra96,ganga96c,hancock96,lineweaver96,rocha96}), or on
maximum likelihood analyses of CMB data sets that directly use model CMB
anisotropy spectra (e.g., \cite{ganga96a},b; \cite{bj96}), can be used to 
constrain parameters, although they do not yet provide strong constraints
in the most interesting regions of parameter space.
Higher precision measurements of CMB anisotropies 
will become available in the next few years, from numerous ground- and 
balloon-based experiments and eventually from the MAP and COBRAS/SAMBA
space observatories.
These measurements can, in principle, constrain cosmological parameters to
unprecedented accuracy. At the same time, our knowledge of galaxy
clustering will improve significantly with the next generation of redshift
surveys, the SDSS and 2dF surveys (\cite{gw95,colless96}). 
(The simulations described here
will be used to generate publicly available mock catalogues with the
geometry and selection criteria appropriate to these surveys.) The ability
to model galaxy formation using cosmological gasdynamic simulations is
improving rapidly, and it may well soon be possible to establish whether 
the apparently complex relation between galaxies and mass implied by
our analysis is a natural outcome of hierarchical clustering in a 
low-$\Omega_0$ universe.

\section*{ACKNOWLEDGMENTS}
We thank Carlton Baugh for providing the APM data shown in 
Figures~\ref{fig:xi} and~\ref{fig:psp} and advice
regarding setting up glass initial conditions. We thank Hugh Couchman
for generously making his AP$^3$M N-body code publicly available and for
helpful discussions on its use.
SMC acknowledges the support of a PPARC Advanced Fellowship.
DHW acknowledges support from NASA Astrophysical Theory
grants NAG5-2882 and NAG5-3111. CSF acknowledges the support of a PPARC 
Senior Fellowship. BR acknowledges support from 
NSF grant EPS-9550487 and matching support from the state of Kansas.
We thank Andrew Liddle for a helpful report.

\end{document}